# Ice templating of porous alumina: supercooling and crystal growth during the initial freezing regime

Audrey Lasalle[♦], Christian Guizard, Jérôme Leloup, Sylvain Deville

Laboratoire de Synthèse et Fonctionnalisations des Céramiques, UMR 3080 CNRS/Saint-Gobain, 84 306 Cavaillon, France

Eric Maire, Agnès Bogner, Catherine Gauthier, Jérôme Adrien, Loïc Courtois

Université de Lyon, INSA-Lyon, MATEIS CNRS UMR5510, F-69621 Villeurbanne, France

**Abstract**

We investigate the ice-templating behaviour of alumina suspensions by in-situ X-rays radiography and tomography. We focus here on the formation and structure of the transitional zone, which takes place during the initials instants of freezing. For many applications, this part is undesirable since the resulting porosity is heterogeneous, in size, morphology and orientation. We investigate the influence of the composition of alumina suspensions on the formation of the transitional zone. Alumina particles are dispersed by three different dispersants, in various quantities, or by chlorhydric acid. We show that the height and the morphology of the transitional zone are determined by the growth of large dendritic ice-crystals growing in a supercooled state, and growing much faster than the cellular freezing front. When the freezing temperature decreases, the degree of supercooling increases. This results in a faster freezing front velocity and increases the dimensions of the transitional zone. It is therefore possible to adjust the dimensions of the transitional zone by changing the composition of alumina suspensions. The counter-ion $Na^+$ has the most dramatic influence on the freezing temperature of suspensions, yielding a predominance of cellular ice crystals instead of the usual lamellar crystals.

---

[♦] email : lasalle.audrey@gmail.com

# 1. Introduction

Ice-templating is a shaping process used to obtain porous materials by freezing a suspension of particles and extracting the ice-crystals by freeze-drying. The potential interest for ceramic material was discovered by Fukasawa in 2001 [1]. This process consists in pouring a ceramic suspension into a mold usually frozen by the bottom. Ice crystals grow unidirectionally according to the thermal gradient, through the suspension, rejecting and packing particles between them. Ice crystals are removed by sublimation to preserve the scaffolds of concentrated particles. A porous green body is thus obtained where the porosity corresponds to the previous ice-crystals. The green body can then be consolidated by sintering [2]. Different pore sizes (from 1 to 100 µm), porosity (from 10 to 90%) and pore morphologies can be obtained. Such materials are currently considered for biomedical application [3], active substance delivery [4], acoustic insulation [5] SOFC [6,7] or piezoelectric applications [8].

Freezing occurs in two stages: an initial transitory regime [9] followed by a steady state [10]. During the initial instants of freezing, the freezing front advances quickly. The microstructure obtained is characterized by two populations of ice crystals: R-crystals (R for randomly oriented) and Z-crystals (for lamellar and oriented ice -crystals). Because of their different orientation, these two populations of crystals do not concentrate ceramic particles with the same efficiency [9]. We have previously shown that the sudden extinction of the R-crystals marks the end of the transitional zone, leaving only lamellar crystals. This corresponds to the beginning of the steady state of the freezing.

For many applications, a continuous porosity is sought. The presence of a graded structure, corresponding to this transitional zone, can be problematic. For instance, in the SOFC design currently investigated by NASA [7], the transitional zone limits the efficiency of the infiltration of the porous electrodes obtained by ice-templating.

Several parameters were identified to have an influence over this transitory regime, such as the particle size [11], the shape of cooling ramp (linear, constant, parabolic) [12] or the application of an electric field (ref: Z. Yumin, H. Luyang, and H. Jiecai, "Preparation of a Dense/Porous BiLayered Ceramic by Applying an Electric Field During Freeze Casting," J. Am. Ceram. Soc., 92[8] 1874-1876 (2009)). In the present paper, we show that the characteristics of this transitory regime are controlled by the initial supercooling conditions. We assess the influence of the composition of alumina suspensions, stabilized with various dispersants. We use X-rays radiography and tomography to respectively observe the freezing front behaviour and characterize the resulting fro-zen microstructure. Although the 1.4 µm resolution does not permit to resolve individual particles, it is high enough to provide an accurate observation of ice-crystals and concentrated particles regions.

## 2. Experimental

### 2.1 Suspensions preparation

Alumina powder (Ceralox SPA 0.5, Sasol, Tucson, AZ, USA), $D_{50} = 0.3$ µm, with a specific surface area of $8m^2.g^{-1}$, is dispersed in distilled water with an organic dispersant or with chlorhydric acid. Alumina content is always $32_{vol}\%$, unless specified. Three sorts of suspensions are prepared. Some contains an ammonium polymethacrylate as dispersant (Darvan CN, Vanderbilt, Norwalk, CT, USA), others contains a sodium polymethacrylate (Darvan 7 Ns, Vanderbilt, Norwalk, CT, USA), and the third type of suspension is dispersed by an ammonium polyacrylate (Darvan 821A, Vanderbilt, Norwalk, CT, USA). These organic dispersants are respectively referred in text as $D[NH_4^+]$, $D[Na^+]$ and $d[NH_4^+]$. They are introduced with a quantity varying from 0.2 to $2_{wt}\%$. Suspensions prepared with chlorhydric acid (HCl) contain $3.1\ 10^{-2}$ or $6\ 10^{-2}$ mol.$L^{-1}$ of HCl. Dispersant (or acid) is stirred with distilled water during 30 min and then the alumina powder is added. Alumina suspensions are ball-milled during 40 h, de-aired before being ice-templated. The dispersant quantity is expressed in weight percent related to the mass of dried powder. The ice-templated alumina suspensions are summarized in the table 1.

### 2.2 Freezing set-up

Suspensions are introduced into a polypropylene mold of 3 mm of diameter, with a syringe. A particular attention is paid to not introduce air-bubbles in suspension. The mold is on a cooper finger frozen by the bottom by a nitrogen liquid flux pumped into a Dewar. The cooling rate is imposed by the liquid nitrogen flow and the temperature profile is registered during experiment, thanks to a thermocouple located near the copper finger surface. The set-up of freezing was previously described [9]. Our measurements have shown that alumina suspensions are frozen at a cooling rate of 3 to 5.3 °C.min$^{-1}$.

### 2.3 Freezing temperature measurement

Two techniques provide the freezing temperature. Cold-Differential Scanning Calorimetry (DSC Q100, TA Instruments, USA) was used on a constant 15 µL volume, introduced into a capsule. The capsule is cooled at 2 or 5°C/min. We consider that the freezing temperature corresponds to the beginning of the endothermic peak and not to the temperature corresponding to the minima reached by the heat flux because a looping with variable amplitude appears at each run. This loop is due to the very important latent heat coming from water to ice transformation. We also use a thermocouple located at the surface of the copper finger using during the freezing experiment. When the radiography of the initial zone is acquired, the freezing temperature corresponds to the temperature when the first ice crystals are observed.

## 2.4 X-ray radiography and tomography

A monochromatic high coherent X-rays beam with an energy of 20.5 keV is sent through the sample. A CDD camera with 2048 x2048 sensitive elements is placed at 20 mm behind the sample. The advancement of the freezing front is followed by radiography with a spatial resolution of 2.8 µm. A sequence of pictures is acquired without moving the sample to follow the advancement of the freezing front, so that we can accurately measure the freezing front velocity. On radiography, the grey level is proportional to the transmission of x-rays, so low values of grey levels correspond to a zone with low absorption of X-rays, that is, ice in our case. High grey levels correspond to the particle-rich phase.

Tomography consists in taking radiographies at different viewing angles on 180°. To ensure a good observation, microstructure must not evolve during acquisition. If we would perform a tomography during the freezing, as the scan takes about ten minutes, the freezing front would move and the microstructure would evolve. We therefore observe the microstructure when the sample is completely frozen. The frozen sample is maintained at a constant temperature during the scan. The spatial resolution is 1.4 µm. After a computer treatment, a 3D map of the absorption coefficient is obtained. Contrary to radiography, the grey level is proportional to the absorption of X-rays.,Low grey levels correspond to the phase with the higher atomic number, the particle-riche phase in our case.

## 2.5 Image analysis

The height of transitional zone is measured from tomography acquisition. A sequence of 1600 projections of 2048 x 2048 pixels is obtained. Such large files are difficult to handle. We resize the picture to obtain a square of 400 x 400 px with a resolution of 4.7 µm. Then each picture in grey level is transformed in binary pictures by using a segmentation criterion, provided as plug-in in the software Fiji [13]. A scan of 1600 binarised pictures is obtained. By image analysis, we calculate the fraction of ice and the fraction of the particle-rich phase. By reporting the variation of the rich particles phase fraction as a function of position z in sample (z is the freezing direction), we can measure the height of the transitional zone. It is characterized by a rapid decrease of the rich particles phase fraction, followed by an increase [9]. This sudden variation originates from the extinction of R-crystals, which concentrate particles in the z direction, while Z-crystals concentrate them mostly in the xy plane, roughly perpendicular to the temperature gradient.

## 3. Results

The freezing temperatures were found in the range -10 and -20 °C, for a cooling rate included between 2 and 5.3 °C.min$^{-1}$ regardless of the composition and the technique of measurement (figure 1). It appears that the counter-ion Na$^+$ used in dispersant seems to be the only parameters affecting the freezing temperatures; lower freezing temperatures are systematically obtained with this ion. In order to decouple the influence of the dispersant and the influence of the particles, we measured, by cold-DSC only, the freezing temperature of aqueous solutions containing the quantity of dispersant (or chlorhydric acid) in similar quantities the alumina suspensions contain (figure 2). Surprisingly, distilled water freezes between -20 and -25°C, far from the equilibrium temperature of 0 °C, and the addition of dispersant or acid increases the freezing temperature. The freezing temperature measured for distilled water can be explained by the specific freezing conditions in the cold-DSC experiments. A very small amount of ultrapure water is put in a clean capsule, so that little nucleation sites are available. Water can be supercooled before the liquid/solid transition occurs. Surfactants (dispersants) and particles incorporated in water act as nucleation sites [11]. The freezing temperature of water, measured in such conditions, cannot be compared to the freezing temperature of the suspensions. At the freezing temperature, the ice nucleus overcome their critical size and can grow into ice crystals [14]. Solidification can proceed.

By tracking the position of the freezing front by X-ray radiography (figure 3), we obtain the instantaneous freezing front velocity, and report its evolution according to the time for different suspensions, in terms of solid loading, nature of dispersant, and quantity of dispersant or chlorhydric acid. All the points are on the same trajectory, but do not necessarily have the same initial and final interface velocity.

Careful observations of the radiographs allow us to rationalize preliminary observations previously reported [9]. When freezing begins, large ice crystals, with a very dendritic morphology, grow very quickly into the suspension. Their growth velocity is five to fifteen times greater than that of the freezing front appearing later (figure 4). When they stop growing, they partially melt back, with the melting starting from the top.

Reconstruction of the tomography data show that the microstructure and magnitude of the transitional zone depends on the composition of the suspension. For the suspension containing $2_{wt}\%D[NH_4^+]$, we clearly observe the end of transitional zone, characterized by the apparition of a continuous lamellar morphology (figure 5b). A perpendicular cross section shows a typical microstructure previously observed, of R-crystals and Z-crystals (figure 6a). This microstructure is less well defined for the suspension with $0.2\%D[NH_4^+]$, but the location of the end of the transitional zone can still be identified. Suspensions containing 0.2 or $2_{wt}\%$ $D[Na^+]$ (figure 5c,d) have a transitional zone larger than that of suspensions containing $D[NH_4^+]$ (figure 5a,b). A perpendicular cross section shows that the transitional zone is just constituted by cellular ice-crystals. No lamellar ice-crystals are observed (figure 6b).

## 4. Discussion

To understand the role of the large dendritic ice crystals, their dimensions, assessed on the radiography, can be compared with the frozen microstructure of the corresponding zone (figure 7.a, b). We observe that the location where these large ice crystals stop coincides with the location of the end of the transition zone. When the transition zone extend beyond the observation window, in the case of suspensions containing $D[Na^+]$, the top of the large dendritic crystals is not observable on the radiographs. It is therefore safe to assume that the growth and dimensions of these crystals corresponds to the formation and extent of the transitional zone, and that these large crystals correspond to the previously identified R-crystals. The large dendritic ice crystals should be the first formed nuclei reaching their critical size. These nuclei are formed in a highly supercooled suspension. The location where their growth come to a halt should thus corresponds to the location where the suspension is at its equilibrium temperature. Similar conclusions were recently reached by Spannuth & al. [15], using a very different approach. They observed that the transitional zone (referred as stage I in their paper) takes end when the sample is warmed to the melting temperature. We can also assume that the top of the large dendritic ice crystals melt back because of the important latent heat released during their growth. The extinction of the R-crystals marks the end of the transitional zone [9]. We enter in the steady state of the freezing regime. Finally the dendritic aspect of the R-crystals can be explained by their very high growth velocity, cellular and dendritic morphology being favoured by high interface velocity.

We observe a different freezing behaviour for the suspensions containing $D[Na^+]$. The presence of $Na^+$, introduced as a counter ion in the dispersant, decreases the freezing temperature. The ammonium polyacrylate is known to exhibit an antifreeze effect [16]. As the sodium polymethacrylate has a similar molecular configuration, we suppose it can exhibit a similar antifreeze activity. This activity is reinforced by the structuring nature of the sodium cation and its capacity to rearrange molecular water [17] and might thus explain the decrease of the freezing temperature. By decreasing the freezing temperature, the supercooling effect increases. As the first ice crystals appearing do so in a supercooled suspension, they grow very fast to quickly reach their equilibrium position, where the freezing temperature and the temperature in the suspension are equal. The greater the degree of supercooling, the greater the freezing front velocity. We know that the freezing front velocity determines the ice crystal morphology, which is lamellar at intermediate velocity and cellular when velocity increases (ref: Qian L, Ahmed A, Foster A, Rannard SP, Cooper AI, Zhang H. Systematic tuning of pore morphologies and pore volumes in macroporous materials by freezing. Journal Of Materials Chemistry 2009;19:5212.). If we report the evolution of the time it takes for the freezing front to propagate through the observation window on the radiography (transit time) as a function of the freezing temperature, we can observe that the greater the degree of supercooling (the lower the freezing temperature), the greater the freezing front velocity (low transit time) (figure 8). The cellular microstructure (figure 6b) in case of D[$Na^+$] might thus be explained by an increase of the freezing front velocity, due to an important supercooling effect.

We can adjust, to some extent, the dimensions of the transitional zone with initial degree of supercooling (figure 9), which increases when the freezing temperature of alumina suspensions decreases. The closer the freezing temperature from equilibrium temperature, the lower the initial degree of supercooling. The position where the suspension is at equilibrium temperature will be closer to the base of the sample and R-crystals will stop growing earlier, resulting in a shorter transitional zone. With these observations, we show that is possible to adjust the height of the transitional zone through the control of the initial supercooling, which is dependent on the composition of the suspension, but also the nucleation conditions, controlled by, among other parameters, the particle size and the shape of the cooling ramp applied. By changing the counter ion of the dispersant used from $NH_4^+$ to $Na^+$, the freezing temperature can be decreased and consequently the height of transitional zone can be increased.

## 5. Conclusions

The current results allow us to rationalize all previous observations of the transitory stage occurring during the solidification of colloidal suspension. The cooling conditions and suspensions composition and characteristics will determine the degree of supercooling reached when the first ice nuclei reach their critical size. The transitory stage is determined by this supercooling, and can be controlled to some extent by adjusting the freezing temperature of the suspension or controlling the nucleation conditions. The later can be achieved by the incorporation of ice nucleation sites, or cooling conditions less favourable to supercooling. These parameters can be used to control the graded structure of the transitional zone, undesirable for applications of porous materials obtained by ice-templating.

## Acknowledgments

Financial support was provided by the National Research Agency (ANR), project NACRE in the non-thematic BLANC programme, reference BLAN07-2_192446. Beamline access was provided by the ERSF, under the proposal MA997. Acknowledgements are due, as usual, to local staff of the beamline: Elodie Boller, Paul Tafforeau and Jean-Paul Valade for the technical and scientific support on the beamline ID-19 at ERSF.

| Alumina content | Nature of additive | Quantity of additive |
|---|---|---|
| $32_{vol}\%$ | $D[NH_4^+]$ | $0.2-0.4-0.7-1-2_{wt}\%$ |
| $50_{vol}\%$ | $D[NH_4^+]$ | $0.2-2_{wt}\%$ |
| $32_{vol}\%$ | $D[Na^+]$ | $0.2-2_{wt}\%$ |
| $32_{vol}\%$ | $d[NH_4^+]$ | $0.2-1_{wt}\%$ |
| $32_{vol}\%$ | HCl | $3.1\ 10^{-2} - 6\ 10^{-2}\ mol.L^{-1}$ |

Table 1. Composition of the ice-templated alumina suspensions

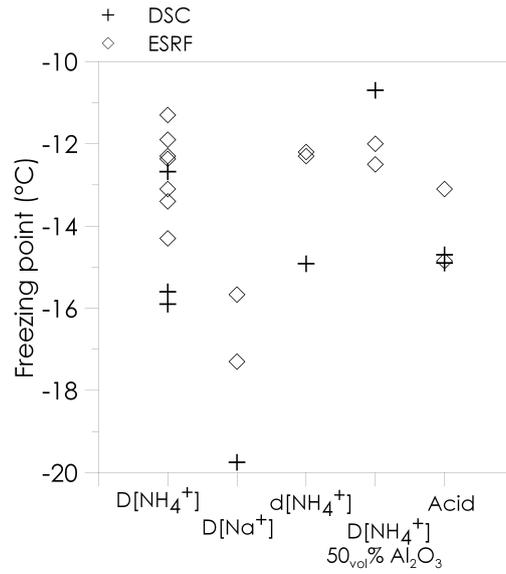

Fig.1. Freezing temperature of alumina suspensions, measured by cold DSC (cross points) and by a thermocouple during ERSF experiments (rhomb points). The freezing temperature is measured for different compositions, the dispersant used is reported on the abscissa and all alumina suspensions contain $32_{vol}\%$ of particles except when it is indicated.

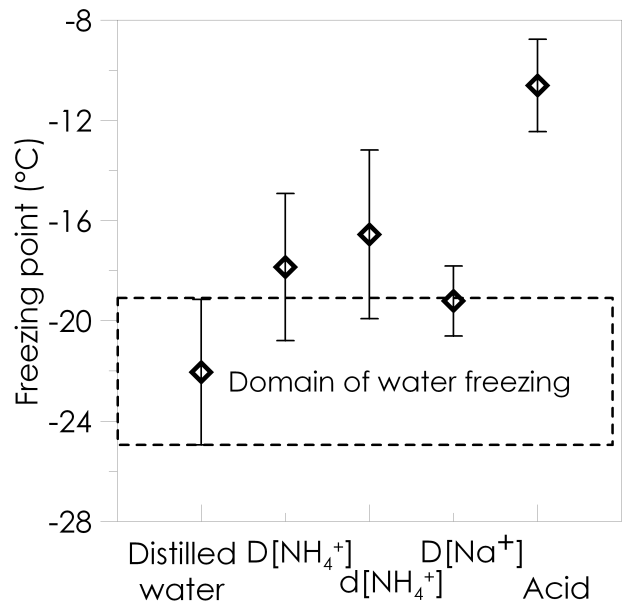

Fig.2. Freezing temperature obtained by cold-DSC, of aqueous solutions (without alumina particles) containing the same dispersant or acid in similar quantity that alumina suspensions contain. The domain where the distilled water freezes is indicated by a dashed rectangle. The cooling rate is 2°C.min$^{-1}$

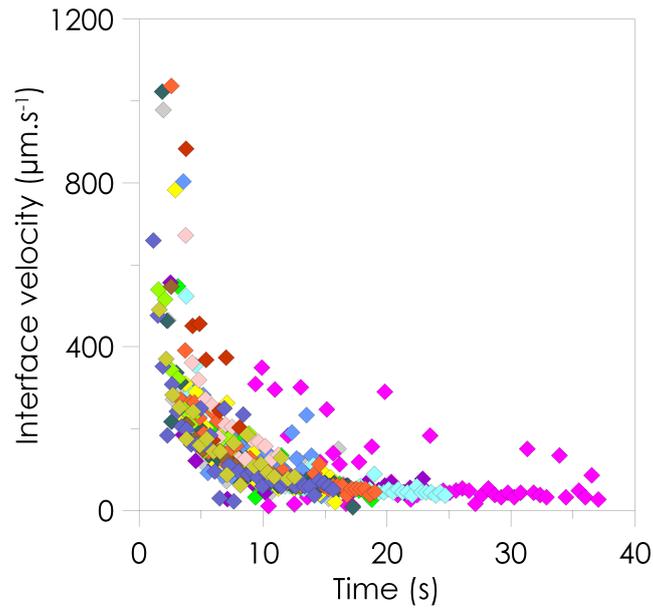

Fig.3. Freezing front velocity versus time measured on X-rays radiographies by picture analysis. Each color corresponds to a composition of an alumina suspension. The composition vary with the alumina content (32 or 50vol%) with dispersant nature (D[$NH_4^+$], D[$Na^+$], d[$NH_4^+$] or HCl) and with dispersant quantity (from 2 to $2_{wt}$%) and from $3.10^{-2}$ to $6.10^{-2}$ mol.$L^{-1}$ for HCl.

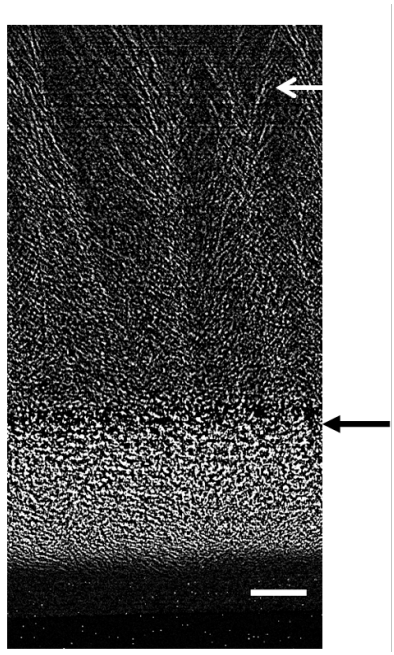

Fig 4. Binary image obtained by subtraction of three consecutive radiographies. The white arrow designates the large ice crystals with a dendritic morphology, the black arrow indicates the freezing front. Scale bar: 200 µm.

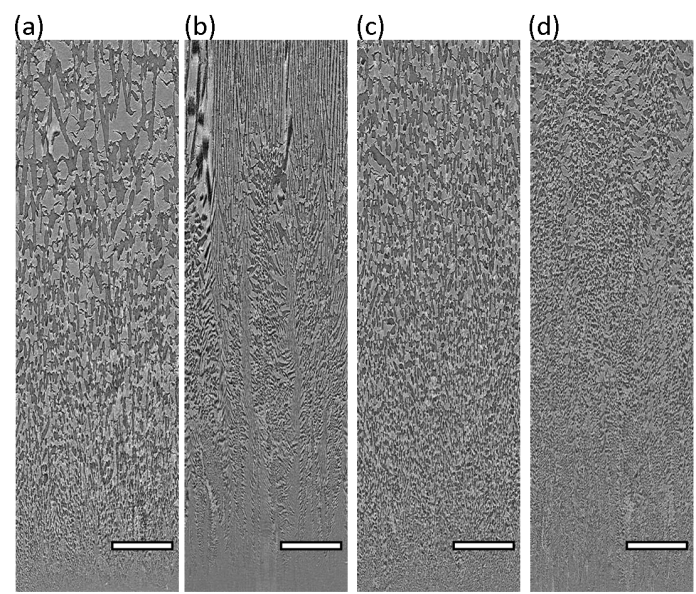

Fig 5. Cross sections of reconstructed tomography, parallel to the freezing direction, showing the frozen microstructure of the transitional zone for the alumina suspensions containing $0.2_{wt}\%D[NH_4^+]$ (a), $2_{wt}\%D[NH_4^+]$ (b), $0.2_{wt}\%D[Na^+]$ (c), $2_{wt}\%D[Na^+]$ (d). The end of the transitional zone can only be observed for the both first suspensions. Scale bars = 200 µm.

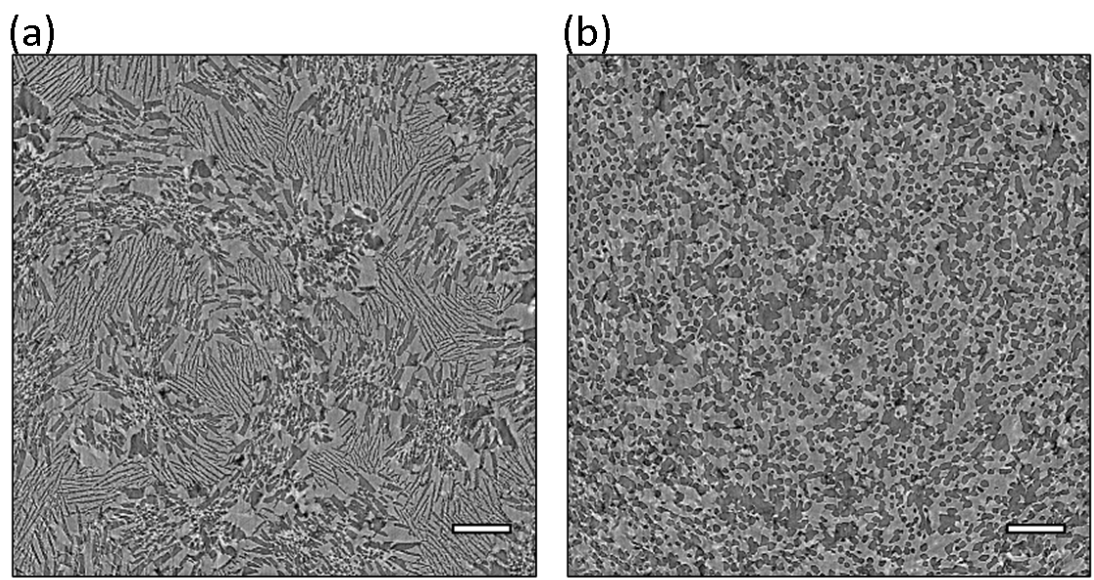

Fig 6. Cross sections of reconstructed tomography, perpendicular to the freezing direction, showing the frozen microstructure of the suspension $2_{wt}\%$ $D[NH_4^+]$ (a) and $0.2_{wt}\%D[Na^+]$ in the transitional zone. The first microstructure is a mix between R and Z-crystals (a) whereas with $D[Na^+]$ no Z-crystals are observed, only cellular R-crystals with a cellular morphology appear. Scale bar = 200 µm

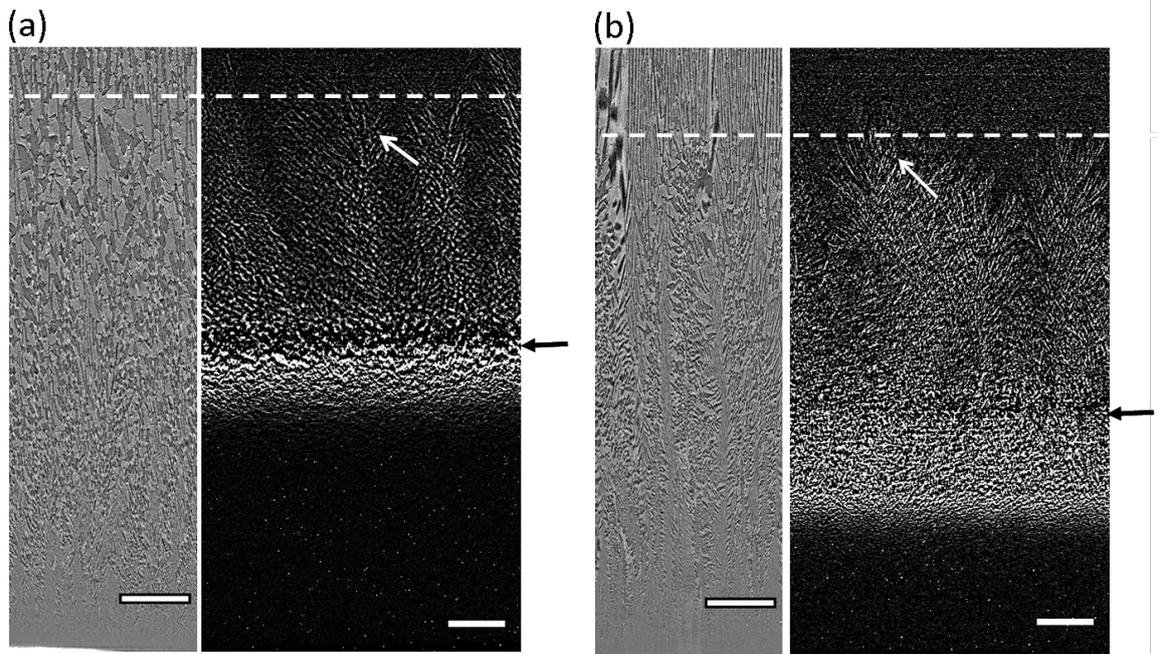

Fig. 7. Binary radiography obtained by subtraction of three consecutive radiographies (right hand side of each image) versus the corresponding reconstructed tomography (left hand side). The dashed line indicates the location where the large dendritic ice crystals (white arrow) stop growing. The black arrows indicates the location of the freezing front. By extending the dashed line on the reconstructed tomography, we observe the large ice crystals stop around the end of the transitional zone. The ice-templated suspensions contain $0.4_{wt}\%D[NH_4^+]$ (a) and $2_{wt}\%D[NH_4^+]$ with $32_{vol}\%Al_2O_3$ (b). Scale bars = 200μm.

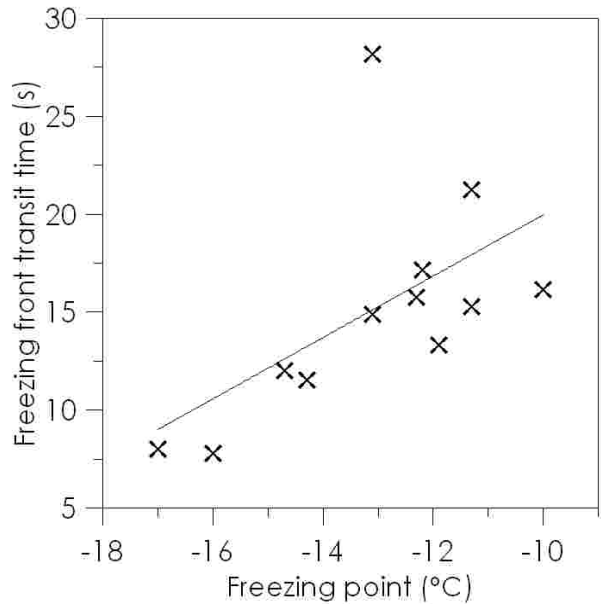

Fig 8. Freezing front transit time for different suspension as a function of the freezing point measured with a thermocouple. The lower values correspond to suspensions prepared with D[Na$^+$]. The greater the supercooling, the higher the freezing front velocity, which is inversely proportional to the freezing front transit time.

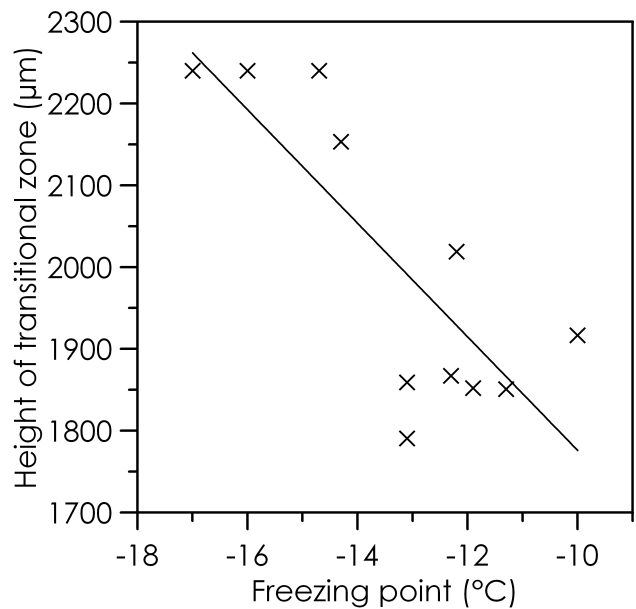

Fig.9. Height of transitional zone versus freezing temperature. The lower the freezing temperature, the higher the transitional zone.